\documentclass[twocolumn,english,groupedaddress,superscriptaddress,nofootinbib]{revtex4-1}
\usepackage{amsfonts}
\usepackage{amsmath}
\usepackage[dvips]{graphicx}
\usepackage{color}
\usepackage{graphicx}
\usepackage{amssymb}
\usepackage{hyperref}
\usepackage{color}
\usepackage{amsthm,amsmath,amssymb}
\usepackage{mathrsfs}
\usepackage{braket}
\usepackage{epstopdf}
\usepackage{dsfont}
\usepackage{setspace}
\usepackage{multirow}
\usepackage{booktabs}
\usepackage{mathtools}
\usepackage{tabularx}
\usepackage{adjustbox}
\usepackage{colortbl}
\usepackage{hhline}
\usepackage{xcolor}
\usepackage{babel}
\usepackage{float}
\usepackage[normalem]{ulem}
\allowdisplaybreaks[4]

\makeatletter
\definecolor{newtxtcolor1}{rgb}{0.5,0,0.5}
\definecolor{newtxtcolor2}{rgb}{0.72,0.46,0.83}
\definecolor{newtxtcolor3}{rgb}{0.86,0.47,0.3}
\definecolor{nblue}{rgb}{0.3,0.3,1.0}
\definecolor{ngreen}{rgb}{0.2,0.7,0.2}
\definecolor{nred}{rgb}{0.9,0.1,0}
\definecolor{nblack}{rgb}{0,0,0}

\definecolor{newtxtcolort}{rgb}{0.8, 0, 0.2}

\begin{document}
	
\title{Foundation Model for Unified Characterization of Optical Quantum States}
\author{Xiaoting~Gao}
\address{State Key Laboratory for Mesoscopic Physics, School of Physics, Frontiers Science Center for Nano-optoelectronics, $\&$ Collaborative Innovation Center of Quantum Matter, Peking University, Beijing 100871, China}

\author{Yan~Zhu}
\address{QICI Quantum Information and Computation Initiative, Department of Computer Science,
The University of Hong Kong, Pokfulam Road, Hong Kong}

\author{Feng-Xiao~Sun}
\address{State Key Laboratory of Information Photonics and Optical Communications and School of Physical Science and Technology, Beijing University of Posts and Telecommunications, Beijing 100876, China}
\address{State Key Laboratory for Mesoscopic Physics, School of Physics, Frontiers Science Center for Nano-optoelectronics, $\&$ Collaborative Innovation Center of Quantum Matter, Peking University, Beijing 100871, China}

\author{Ya-Dong~Wu}
\email{wuyadong301@sjtu.edu.cn}
\address{John Hopcroft Center for Computer Science, Shanghai Jiao Tong University, Shanghai 200240, China}

\author{Qiongyi~He}
\email{qiongyihe@pku.edu.cn}
\address{State Key Laboratory for Mesoscopic Physics, School of Physics, Frontiers Science Center for Nano-optoelectronics, $\&$ Collaborative Innovation Center of Quantum Matter, Peking University, Beijing 100871, China}
\address{Collaborative Innovation Center of Extreme Optics, Shanxi University, Taiyuan, Shanxi 030006, China}
\address{Peking University Yangtze Delta Institute of Optoelectronics, Nantong, Jiangsu 226010, China}
\address{Hefei National Laboratory, Hefei 230088, China}
    
\begin{abstract}
Machine learning methods have been used to infer specific properties of limited families of optical quantum states, but a unified model that predicts a broad range of properties for practically relevant—especially multimode non-Gaussian—states without full tomography is still lacking. Here we introduce the first foundation model for the characterization of optical quantum states across a wide range of complexity, defined by three key factors: non-Gaussianity, number of modes, and degree of squeezing.
We show that a single model pretrained on low-complexity states can be directly applied to characterize states of higher complexity. With limited fine-tuning, the model adapts to downstream tasks such as predicting quantum fidelity and Wigner negativity over a broad class of experimentally relevant states, including strongly non-Gaussian Schrödinger cat states, multimode systems with up to ten modes, and highly squeezed states with squeezing levels up to $10.4$dB. Our results establish a unified framework for characterizing optical quantum states from limited measurement data, enabling efficient certification of quantum states relevant to optical quantum information computation, communication and metrology.
\end{abstract}
\maketitle
\section{Introduction}
Optical quantum systems provide an important physical platform for quantum information processing. 
As the field moves toward universal quantum computation, the focus has recently shifted from small-scale demonstrations to scalable, classical intractable ones, namely multimode non-Gaussian states~\cite{Walschaers2021nongau}. 
A central challenge in optical quantum information is the experimental characterization of quantum states from measurements, including quantum correlation detection~\cite{Adesso_2007,gessner2017entanglement,21t1-dqn6}, quantum certification~\cite{aolita2015reliable,chabaud2021efficient,chabaud2020building,wu2021efficient}, and full quantum state tomography~\cite{lvovsky2009continuous,PhysRevLett.120.090501,he2024efficient}. Recent advances in deep learning have opened new avenues for learning quantum systems from quantum measurement data~\cite{du2025artificial}. These methods enable accurate predictions of quantum properties of optical states~\cite{cimini2020neural,PhysRevLett.130.210601,gao2023correlation}, and in some cases, full state reconstruction~\cite{tiunov2020experimental,PhysRevLett.127.140502,ahmed2021classification,fedotova2022continuous}.

Despite this progress, both the intrinsic hardness of classical simulation and the difficulty of experimentally preparing scalable non-Gaussian states severely constrains the availability of quantum data required for training a deep learning model.
Moreover, existing machine learning models are typically optimized for task-specific objectives, which often confines them to interpolating within in-distribution (ID) data of comparable complexity~\cite{cimini2020neural,PhysRevLett.127.130503,gao2023correlation}.
Overcoming these limitations requires deep neural networks equipped with transfer learning capabilities that enable out-of-distribution (OOD) generalization beyond the training regime.
Recent work suggests that neural network models for learning quantum systems can generalize across systems of different sizes~\cite{wang2022,wu2024learning}, exemplified by the successful extrapolation of quantum fidelity estimation for multi-qubit systems~\cite{shaw2024}.
These progresses raises a natural question: can a model trained on small-scale, weakly non-Gaussian, low-photon-number states be effectively adapted to larger systems exhibiting stronger non-Gaussianity and higher photon numbers?

Meanwhile, foundation models, such as GPT, have achieved remarkable success in natural language processing. By pretraining on massive unlabelled datasets and subsequently fine-tuning for specific tasks, this learning paradigm has significantly enhanced generalization across diverse downstream applications. This pretraining–fine-tuning paradigm has also been applied to quantum many-body systems, where neural networks representing ground states are pretrained and then fine-tuned for property prediction~\cite{PhysRevB.107.075147,PhysRevResearch.6.043280,rende2025foundation}. However, such approaches have not been explored for learning optical quantum states.

In this work, we introduce a foundation model for optical quantum state characterization, unifying the prediction of diverse quantum properties within a single neural network architecture. The model is designed following a pretraining–fine-tuning strategy, in which it first learns self-supervised representations of computationally tractable optical states and is then adapted to increasingly complex regimes. Specifically, optical states are categorized by three factors governing their classical simulation complexity—the number of modes $m$, maximal squeezing degree $\xi$ among all modes, and non-Gaussianity quantified by the stellar rank $r$. The model is first pretrained on states of lower complexity (Stage~1) to predict homodyne measurement statistics. It is then evaluated on more challenging states (Stage~2) to assess out-of-distribution generalization without additional training. Finally, it is fine-tuned on limited data from single-mode strongly non-Gaussian states, highly squeezed states, or multimode non-Gaussian states with up to ten modes (Stage~3) to predict key quantum properties, such as Wigner negativity and fidelity. This framework establishes a pathway toward scalable, generalizable machine learning models for optical quantum state characterization, paralleling the role of foundation models in other domains of science.

\begin{figure*}[hbt]
    \centering
    \includegraphics[width=1\linewidth]{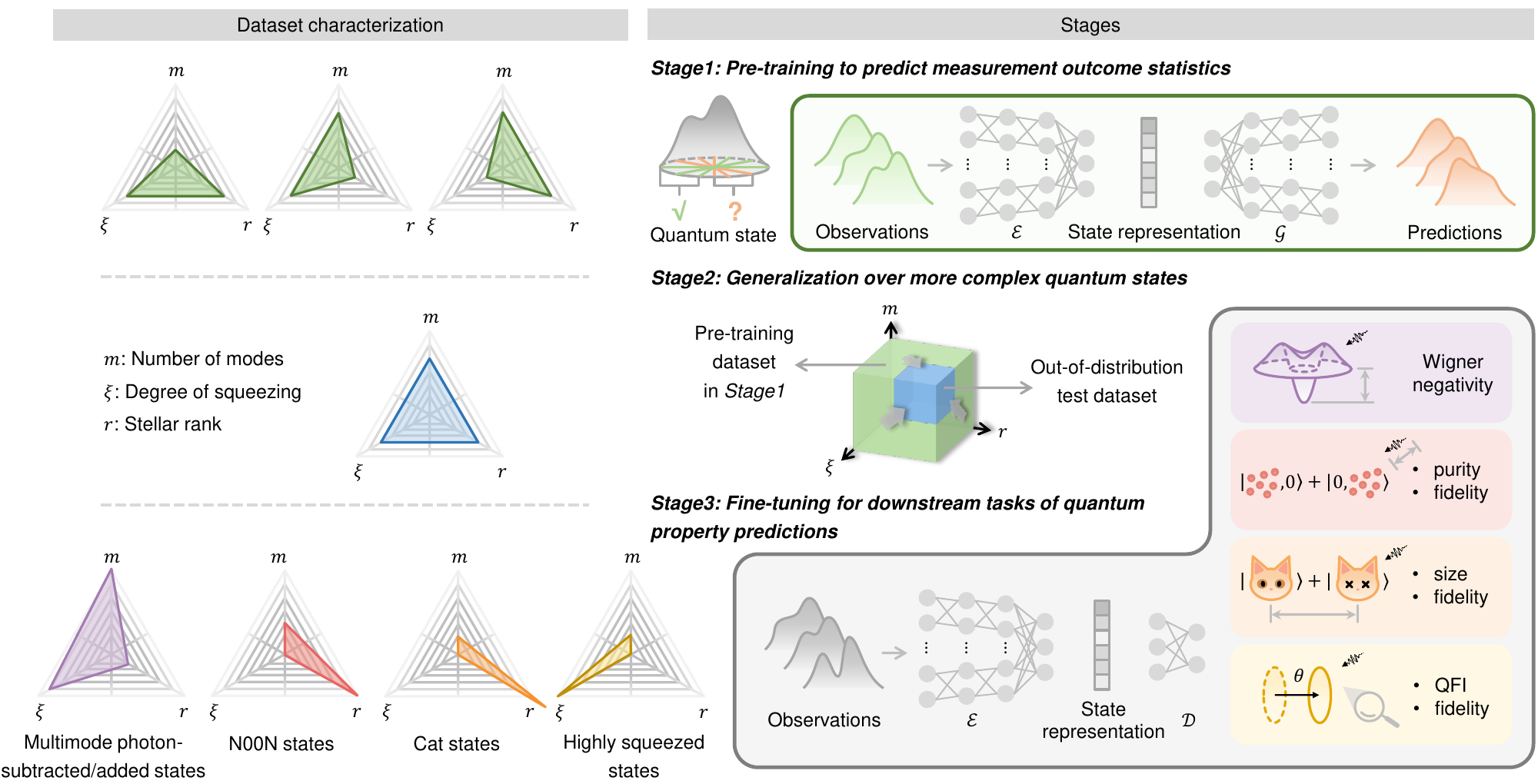}
    \caption{\textbf{Dataset characterization and three-stage framework of transfer learning for optical states.}
    All optical quantum states considered are represented in an $(m,\xi,r)$-space spanned by the number of modes \(m\), the squeezing degree \(\xi\), and the stellar rank \(r\), three axes that correlate with the difficulty of classical simulation.
    In Stage~1, a representation network $\mathcal{E}$ encodes partial measurements and a generation network $\mathcal{G}$ predicts the corresponding outcome statistics. Pretraining is performed on states that are relatively tractable for classical simulation.
    In Stage~2, we test the model on states outside the pretraining distribution, evaluating its ability to generalize the learned representation to more complex quantum states.
    In Stage~3, the pretrained model is fine-tuned for downstream quantum property prediction tasks, focusing on representative families of optical states.}
    \label{fig-1}
\end{figure*}
\section{Results}
\subsection{Optical quantum states construction}
To obtain general optical quantum states in our numerical experiments, we consider them within the stellar formalism~\cite{Chabaud2020stellar,Chabaud2021classical,Chabaud2022holomorphic}. 
In this formalism, quantum states are described by their stellar function (see Methods for details).
The non-Gaussian properties of the states are then described by the distribution of zeros of this function over the complex plane, and the number of zeros defines the stellar rank.
A multimode pure state that admits a decomposition of the form $G|C\rangle$ has finite stellar rank $r\in\mathbb{N}$, where $|C\rangle$ is a core state that has bounded support in the multimode Fock basis, and $G$ is a multimode Gaussian unitary. 
States that do not admit such a decomposition, such as cat states, are assigned $r=+\infty$. 
As Gaussian unitaries preserve $r$, the stellar rank $r$ of a core state fully determines the non-Gaussianity of the associated multimode state. 
In particular, all Gaussian states have $r=0$. 
\par
For pretraining, we first restrict to multimode states with finite stellar rank, i.e., $ G|C(r)\rangle$ with $r\in\mathbb{N}$. 
Using the Bloch–Messiah decomposition, any $m$-mode Gaussian unitary $ G$ can be decomposed as a passive linear transformation followed by a product of $m$ single-mode squeezings, and another passive linear transformation, together with $m$ single-mode displacements, thereby introducing numerous free parameters.
Moreover, optical losses are introduced through a series of single-mode loss channels on the pure state $ G|C(r)\rangle$, yielding an $m$-mode mixed state
\begin{align}
\rho
=\bigotimes_i\mathcal L_i\left(
 G\,|C(r)\rangle\!\langle C(r)|\, G^\dagger
\right)
\label{eq:rho}
\end{align}
where each $\mathcal L_i$ denotes the loss channel on mode $i$ with channel efficiency $\eta_i$~\cite{ou1992realization,LossChannelLect}. 

\subsection{Foundation model for optical quantum state characterization}
Characterizing the optical quantum states, especially non-Gaussian states with complex high-order correlations~\cite{Walschaers2021nongau}, is crucial yet challenging. 
These states live in an infinite-dimensional Hilbert space, which in numerical experiments is often truncated in a Fock basis. 
In the multimode regime, the Hilbert space grows exponentially with the number of modes, further amplifying the difficulty of state characterization. 
Here we introduce a foundation model for optical quantum state characterization (OSFM), which is pretrained to learn compact representations of states from homodyne measurement data and transfers to few-shot downstream quantum property-prediction tasks.
\par
OSFM is organized in three stages, as illustrated in the right panel of Fig.~\ref{fig-1}.
In the first stage, OSFM is pretrained to predict measurement outcome statistics through a generative query network~\cite{zhu2022flexible,GQN_Science} consisting of a representation network $\mathcal E$ and a generation network $\mathcal G$.
In optical settings, an experimenter usually has access to a restricted set of homodyne measurements, indicated by $\mathcal{M}$. 
For each mode $i$, the quadrature operator $ x_{i}(\theta)=1/{\sqrt{2}}\!\left( a_i\,e^{-i\theta} +  a_i^{\dagger} e^{i\theta}\right)$ at phase $\theta$ defines a projective measurement with continuous outcome $x\in\mathbb R$ and marginal distribution
$p^{(i)}(x|\theta)=\langle x_{i}(\theta)\vert\,\rho^{(i)}\,\vert x_{i}(\theta)\rangle$, where $\rho^{(i)}$ is the reduced state of mode $i$. 
Specifically, we sample 100 phases uniformly distributed within $[0,\pi)$ for each mode $i\in\{1,\ldots,m\}$ as the measurement set $\mathcal{M}$. 
The combination of mode index and measurement phase $(i,\theta)$ thus defines an arbitrary local homodyne measurement, denoted by $M\in\mathcal{M}$, and each corresponding marginal $p^{(i)}(x|\theta)$ is discretized into a 50-bin histogram $P$. 
\par
As shown in Stage~1 of Fig.~\ref{fig-1}, for a given optical state, OSFM uses the known marginals $\{P_j\}_{j=1}^{s}$ measured at a random subset $\mathcal S=\{M_j\}_{j=1}^{s}\subset\mathcal M$ as the input context.
$s$ denotes the number of measurements in subset $\mathcal{S}$.
With the randomly selected pairs $\{M_j,P_j\}_{j=1}^s$ from unknown state $\rho$, OSFM produces an average state representation $\mathbf{z}\coloneqq \frac{1}{s}\sum_{j=1}^s\mathcal{E}(M_j,P_j)$ from the representation network $\mathcal{E}$, and then outputs predicted marginals $\{P_j^\prime\}_{j=1}^q$ according to the query set $\mathcal{Q}=\{M_j\}_{j=1}^q$ from the generation network $\mathcal{G}$.
\par
\begin{figure*}[hbt]
    \centering
    \includegraphics[width=1\linewidth]{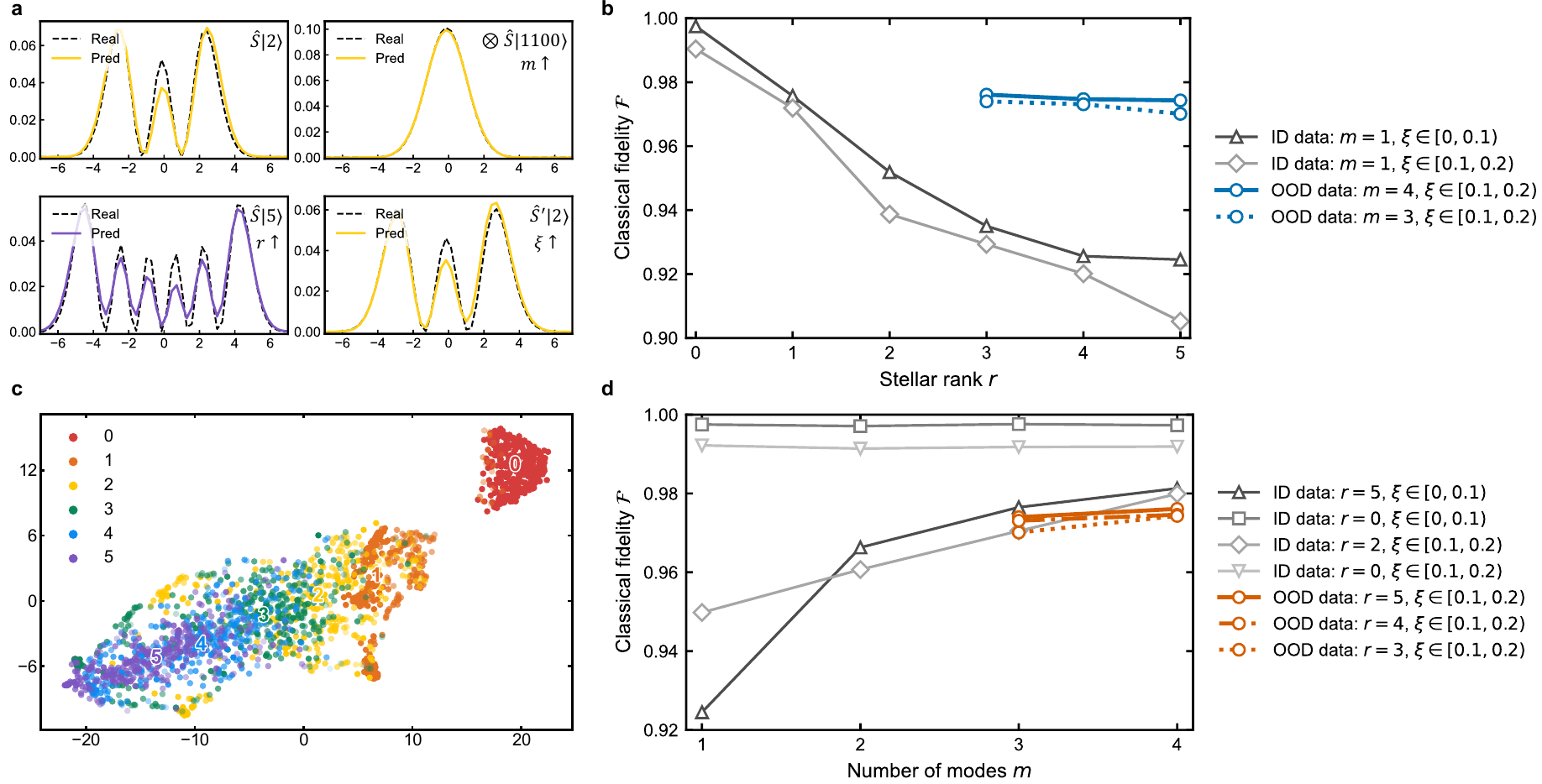}
    \caption{
    \textbf{Representation learning and prediction on homodyne measurement statistics.}
    \textbf{a}, Comparison between the predicted and true homodyne marginal distributions for test states with different parameters. 
    \textbf{b}, Classical fidelity $\mathcal{F}$ decreases with stellar rank $r$ for both in-distribution (ID; within the pretraining distribution) and out-of-distribution (OOD; outside it) data.
    \textbf{c}, 2D embeddings of state representation vectors colored by stellar rank, showing smooth hierarchical organization in the representation space. 
    \textbf{d}, Classical fidelity $\mathcal{F}$ versus the number of modes $m$. 
    For Gaussian states ($r=0$), $\mathcal{F}$ remains high and essentially unchanged across $m$. 
    For non-Gaussian states ($r>0$), $\mathcal{F}$ increases with $m$ and tends to plateau.
    }
    \label{fig-2}
\end{figure*}
The pretraining process minimizes a loss that penalizes the difference between the predicted marginals $\{P_j^\prime\}_{j=1}^q$ and the true marginals $\{P_j\}_{j=1}^q$. 
To further shape the representation space, we add a weighted triplet loss term~\cite{Schroff_2015_CVPR}. 
This additional penalization tends to shorten the distance between representations obtained from the same state, whereas enlarge the distance between those from different states. 
Hence, representation vectors derived from different marginals of the same state stay close in representation space and remain far from those of other states.
\par
Based on the aforementioned optical quantum state construction, we parametrize the states by three factors $(m,\xi,r)$, where $m$ is the number of modes, $\xi$ denotes the maximal squeezing amplitude, and $r$ is the stellar rank that induces a hierarchy of non-Gaussianity. 
Typical families of optical states occupy distinct regions of this \((m,\xi,r)\) space (left panel of Fig.~\ref{fig-1}).    
In the pretraining process, OSFM are fed with states sampled broadly from the parameter space shown as a green hull in Stage~2 in Fig.~\ref{fig-1}. 
Generally, growth along $(m,\xi,r)$ enlarges the effective dimension of Hilbert space, thereby increasing the burden of classical simulation.
After pretraining, we assess zero-shot generalization on OOD states in the blue corner outside the pretraining hull in \((m,\xi,r)\)-space.
OSFM remains accurate in predicting their homodyne measurement distributions without resorting to more demanding classical simulations required for such states.
Stage~3 uses the pretrained representation network \(\mathcal E\) to produce state representations for downstream tasks, with lightweight network $\{\mathcal{D}_i\}$ fine-tuned to predict key quantum properties such as Wigner negativity~\cite{Anatole_Kenfack_2004,Walschaers2021nongau,cimini2020neural} and quantum Fisher information (QFI)~\cite{RevModPhys.90.035005,MATTEO_QFI}, focusing on typical Gaussian and non-Gaussian families with optical loss, spanning a far broader region in the parameter space. 
\par
Taken together, these stages show that the compact representation learned in Stage~1 enables robust generalization on OOD states and efficient transfer across diverse quantum prediction tasks, as we discuss in the following paragraphs.

\subsection{Pretraining and out-of-distribution generalization}
For pretraining, we generate $6\,000$ simulated multimode optical states using Eq.~(\ref{eq:rho}) with randomly sampled physical parameters. 
The possible measurement set \(\mathcal M\) consists of \(100\) local homodyne measurements per mode, with phases uniformly spaced over $[0,\pi)$.
For each state, we sample $10$ to $15$ measurements per mode to form the context subset \(\mathcal S=\{M_j\}_{j=1}^{s}\); the remaining measurements form the query subset \(\mathcal Q=\mathcal M\setminus\mathcal S\). 
Given \(\mathcal S\), OSFM encodes the observations into a $32$-dimensional state representation vector $\mathbf z$ and then predicts the outcome distribution conditioned on any query measurement \(M_j\in\mathcal Q\).
The performance is quantified by the classical fidelity $\mathcal{F}$ between OSFM's predictions and the true distributions of queries. 
\par
Examples are illustrated in Fig.~\ref{fig-2}\textbf{a}, demonstrating how the prediction performance varies along each axis of the \((m,\xi,r)\) space.
The top-left panel shows results of a squeezed Fock state $ S|2\rangle$ with squeezing degree $\xi=0.1$, where $S$ denotes the single-mode squeezing operator $S(\xi)=\exp\!\left[\frac{1}{2}\left(\xi^{*}\hat a^{2}-\xi\,\hat a^{\dagger 2}\right)\right]xw$. 
The remaining three panels respectively probe the three axes of the parameter space and increase each parameter to the maximum of the pretraining green hull.
\par
Increasing non-Gaussianity by raising the stellar rank of state $ S|2\rangle$ from \(r=2\) to \(r=5\) (bottom-left) produces a much more complicated marginal, and the prediction task is thus intrinsically harder.
Results shown in Fig.~\ref{fig-2}\textbf{b} also indicate this trend. 
For Gaussian states with $r=0$, OSFM can perfectly learn the tractable Gaussian marginals.
As the marginal structure becomes richer with increasing stellar rank \(r\), the classical fidelity decreases on in-distribution states.
For lossy non-Gaussian states with stellar rank $r=5$ and squeezing $\xi\in[0.1,0.2)$, OSFM can still achieve an average classical fidelity over $0.90$.
\par
We further estimate the generalization of OSFM on the more complex OOD states. 
While classical fidelity still declines as the stellar rank \(r\) increases, the overall fidelity of OOD states is higher than that of ID states with fewer modes. 
This is due to the single-mode marginals of the reduced multimode non-Gaussian state tend to remain Gaussian and are therefore easier to predict, as exemplified in the top-right panel of Fig.~\ref{fig-2}\textbf{a}.
Correspondingly, Fig.~\ref{fig-2}\textbf{d} shows that the classical fidelity with increasing number of modes \(m\) is almost flat for Gaussian states and improves for non-Gaussian states.
Turning to squeezing, increasing \(\xi\) from $0.1$ to $0.2$ for \( S'\lvert2\rangle\) (bottom-right of Fig.~\ref{fig-2}\textbf{a}) slightly reduces fidelity, yet the predicted marginal remains in close agreement with the ground truth.
\par
After being trained on broad and diverse optical quantum states, OSFM can capture generalizable patterns and extract latent representations of states under consideration.
To visualize how these latent representations are organized, we apply t-distributed stochastic neighbor embedding (t-SNE)~\cite{van2008visualizing} algorithm to the learned state representation vectors, yielding a two-dimensional (2D) layout in which nearby points correspond to similar representations.
In Fig.~\ref{fig-2}\textbf{c}, each point represents one optical state, the colour encodes its stellar rank $r$, and the transparency reflects the largest weight of the highest-stellar-rank components in the core state (larger transparency indicates a smaller maximum weight). 
For instance, a state \(c_0\lvert 1,0\rangle + c_1\lvert 3,1\rangle + c_2\lvert 2,2\rangle\) with \(r=4\) is rendered with transparency \(\max\{|c_1|,|c_2|\}\).
In the layout, two clusters emerge obviously. 
Gaussian states with $r=0$ form a compact cluster well separated from all the non-Gaussian states with $r>0$. 
A small number of lightly coloured orange points ($r=1$) are mixed with the Gaussian cluster.
They correspond to states whose non-Gaussian components have very small amplitudes, so their single-mode marginals remain close to Gaussian and the learned representations are pulled toward the $r=0$ cluster.
\par
Within the non-Gaussian cluster, the embedding follows the ordering defined by the stellar hierarchy.
As the stellar rank increases from 1 to 5, neighbouring ranks get closer and the boundaries blur. 
This trend accords with theoretical conclusions that 
(i) states with finite stellar rank can always be approximated arbitrarily well by higher-rank states with respect to the trace norm, and 
(ii) the stellar robustness, e.g., the minimal trace distance gap to lower-rank states, is expected to decrease with the increasing stellar rank~\cite{Chabaud2020stellar}. 
These properties of non-Gaussian states provide a natural explanation for the unclear boundaries and closer spacing as $r$ grows in the representation space.
The embedding shows that OSFM’s compact representation not only well separates the Gaussian and non-Gaussian boundary, and also organizes non-Gaussian states along the known stellar hierarchy.
\par
To summarize, after pretraining on the computationally tractable region, OSFM extrapolates to more complex states and preserves accurate predictions without resorting to additional large-truncation simulations.

 \begin{figure*}[hbt]
    \centering
    \includegraphics[width=1\linewidth]{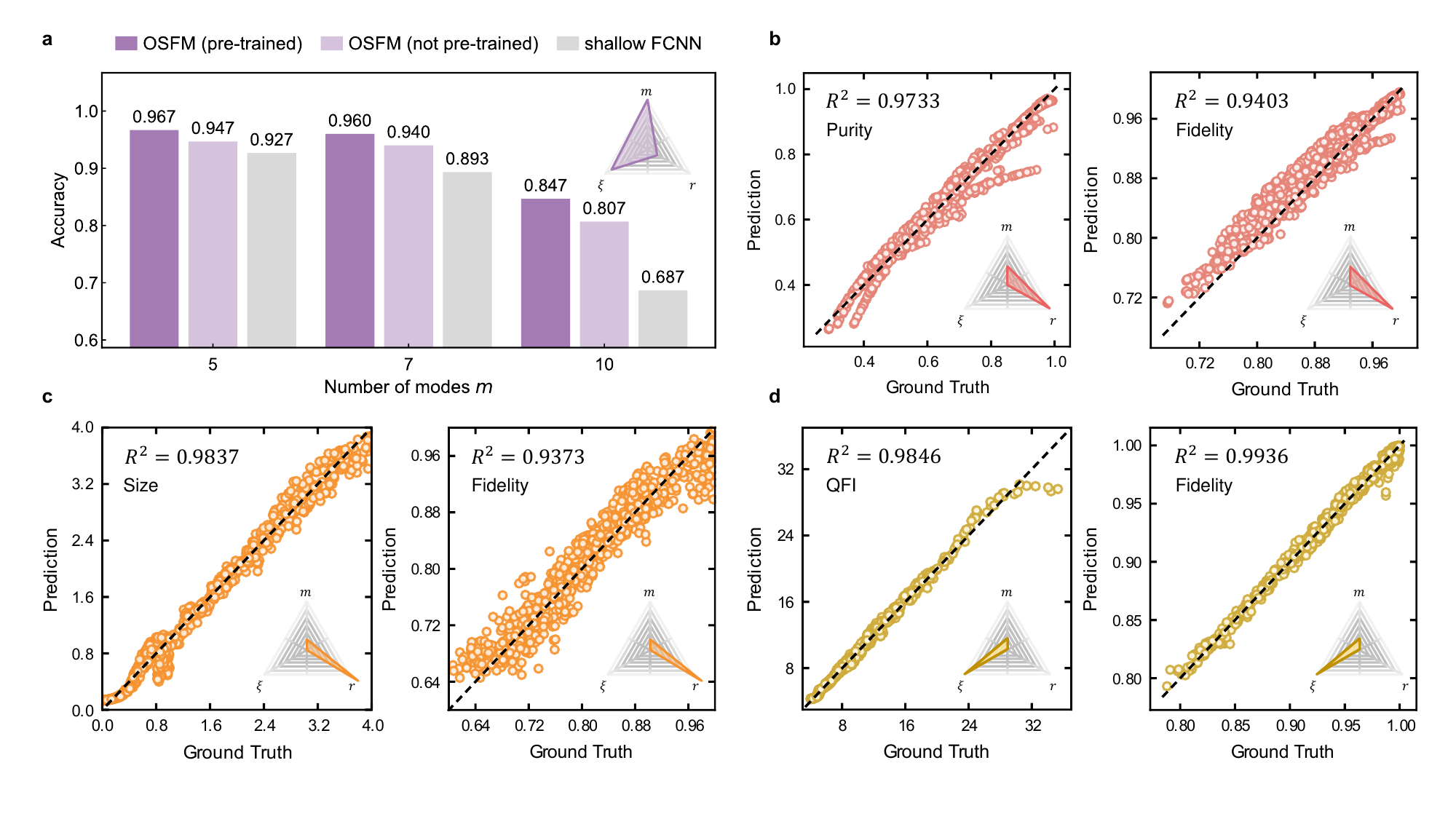}
    \caption{
    \textbf{Fine-tuning for downstream quantum property prediction tasks on state families that lie far outside the pretraining distribution.} 
    \textbf{a}, Comparison of prediction accuracies on Wigner negativity for OSFM with and without pretraining, and for a shallow feedforward neural network baseline, across multimode photon-subtracted/added states with $m=5,7,$ and $10$. 
    \textbf{b}, Predictions of purity and fidelity for N00N states, a non-Gaussian family with finite stellar rank exceeding the pretraining range. 
    \textbf{c}, Predictions of size and fidelity for Schr\"odinger cat states, which possess infinite stellar rank. 
    \textbf{d}, Predictions of quantum Fisher information (QFI) and fidelity for highly squeezed states.}
    \label{fig-3}
\end{figure*}
\subsection{Few-shot downstream transfer}
Having trained OSFM to learn a compact state representation, we now estimate whether this representation can be transferred to specific quantum tasks in the few–shot regime. 
In the pretraining stage, we construct optical states based on a general decomposition form Eq.~\ref{eq:rho} with relatively small physical parameters.
For the fine-tuning task, we now consider state families that lie far outside the pretraining hull in the $(m,\xi,r)$-parameter space, extending toward the large-$m$ (multimode), large-$\xi$ (highly squeezed), or large-$r$ (N00N, cat) ends of the axes. 
For each family, we extract state representations from the pretrained representation network $\mathcal E$ and perform downstream tasks by applying prediction networks $\{\mathcal D_i\}$, fine-tuned on small, task-specific datasets, to these representations. 
\par
This is of great importance in the quantum domain, especially for many-body systems, where large labelled datasets are often scarce owing to the prohibitive cost of classical simulation, which scales exponentially with the Hilbert-space dimension.
As shown below, OSFM trained from Stage~1 transfers well across state families even quite far from the pretraining distribution, allowing for reliable property prediction with only a few hundred labelled states.

\subsubsection{Multimode photon-subtracted/added states}
Multimode continuous-variable quantum optics provides a powerful platform for a multitude of quantum applications. 
As Gaussian resources are efficiently classically simulatable, introducing non-Gaussianity becomes essential, such as photon subtraction or addition ~\cite{subtraction2007science,subtraction-np2009,ra2020non}.
However, full characterization of such states via homodyne tomography is impractical, as the required quadrature settings and classical post-processing scale exponentially with the number of modes.
In this context, fine-tuned OSFM offers a practical alternative by inferring key properties directly from a small set of marginals. 
Here we focus on the negativity of the Wigner function, a pivotal non-Gaussian resource linked to quantum advantage~\cite{PhysRevLett.109.230503,Chabaud2023PRL}.
\par
Concretely, we generate multimode photon-subtracted or photon-added states with $m=5,7$, and $10$ modes.
In each case, we first generate \(600\) random \(m\)-mode Gaussian states with Wigner function $W_0$, whose maximal squeezing degree is set to $\xi_\text{max}=8\text{dB}$. 
For two-thirds of $W_0$, we randomly choose between single-photon subtraction and addition on a randomly selected mode.
When only a single photon is subtracted or added, the corresponding Wigner functions $W_{\pm}$ can be obtained analytically~\cite{PhysRevA.96.053835}. 
\par
We thus produce a mixture of Gaussian states with \(r=0\) and degaussified states with stellar rank \(r=1\), from which we compute the homodyne marginals at three phases \(\theta\in\{0,\pi/3,2\pi/3\}\) for each mode, serving as inputs to the downstream Wigner negativity classifier.
Wigner negativity is assessed via the minimum of the analytical Wigner function. 
A state is labelled negative if \(\min_{\beta} W(\beta)<\tau\), where the threshold \(\tau\) is chosen so that negative and non-negative samples are roughly balanced. 
The downstream classifier is trained with the binary cross-entropy loss.
\par
After fine-tuning, as illustrated in Fig.~\ref{fig-3}\textbf{a}, on datasets with \(m\in\{5,7,10\}\) modes, the pretrained OSFM attains classification accuracies of \(0.967\), \(0.960\), and \(0.847\), respectively.
For comparison, we benchmark against two reference models. 
First, training the representation network of OSFM from a random initialization—without any pretraining—reduces accuracy in every case. 
Second, a commonly used three-layer fully connected neural network (FCNN) baseline performs further behind~\cite{cimini2020neural}, reaching only \(0.927\), \(0.893\), and \(0.687\) for \(m=5,7,10\), respectively. 
The pretrained OSFM significantly outperforms the other two methods, and the gap widens with the number of modes.
\par
Furthermore, pretraining proves more beneficial in the data-scarce regime. 
With only \(75\) labelled states for fine-tuning, the pretrained OSFM achieves accuracies of \(0.929\) for \(m=5\) and \(0.913\) for \(m=7\), whereas the same architecture trained without pretraining reaches \(0.873\) and \(0.876\), and the shallow FCNN yields \(0.827\) and \(0.800\).
These results indicate that the representation learned in Stage~1 can already extract the non-Gaussian structure relevant to Wigner negativity detection, preserving high accuracy even when only a handful of measurements and labels are available.

\subsubsection{N00N states}
We now apply our model to another optical state family with finite stellar rank $r$ extending beyond the pretraining regime, the N00N state~\cite{lee2002quantum}.
In quantum optics, a N00N state is a superposition of $N$ photons across two optical modes
\begin{align}
    \lvert \mathrm{N00N}\rangle=\frac{1}{\sqrt{2}}\left(\lvert N,0\rangle + e^{iN\phi}\lvert 0,N\rangle\right),    
\end{align}
which, in an ideal lossless interferometer, achieves Heisenberg–limited phase sensitivity that scales as $1/N$ and is therefore a central resource for quantum metrology and sensing~\cite{RevModPhys.90.035005}. 
In practice, however, N00N states are notoriously fragile.
The loss of merely a single photon renders this state useless since it collapses into an incoherent mixture of Fock product states, which cannot acquire any phase information.
Accordingly, once loss is present, the metrological advantage no longer improves monotonically with $N$~\cite{Dorner2009PRL,PhysRevLett.130.123603}. 
Given this loss fragility, we evaluate whether OSFM can detect purity and fidelity of lossy N00N states at large $N$, using only limited homodyne marginals.
\par
Transfer learning is performed by fine-tuning the representation network of OSFM together with a linear prediction head, optimized with the mean square error (MSE) loss.
We sample $600$ lossy N00N states with photon numbers $1\leq N\leq8$ to form the tuning dataset, reaching stellar rank up to $r=8$, beyond the pretraining range.
$10$ to $15$ marginals per mode are randomly selected as input.
For each state, we compute the purity and the fidelity between the ideal pure N00N state and its lossy counterpart.
\par
Fig.~\ref{fig-3}\textbf{b} compares OSFM's predictions with the ground truth of the test states. 
Averaged over all test states, the coefficients of determination are $R^2 = 0.9733$ for purity predictions and $R^2 = 0.9403$ for fidelity predictions.
These results show that the representation network learned by OSFM transfers robustly to high-$N$ N00N states in the few-shot regime.

\begin{figure*}
    \centering
    \includegraphics[width=1.0\linewidth]{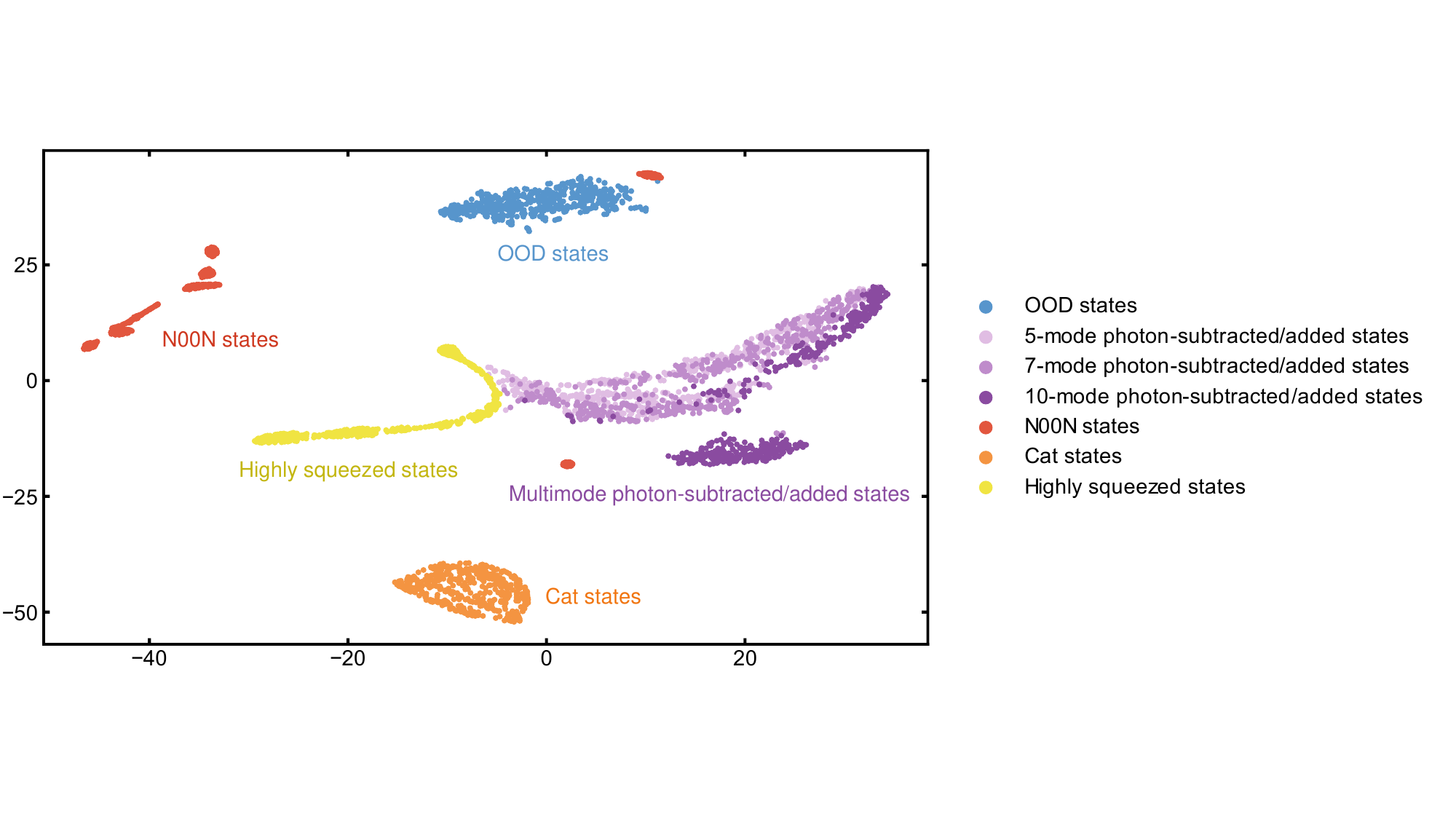}
    \caption{\textbf{The 2D embeddings of diverse optical states using t-SNE.}
    The figure shows the 2D embeddings of state representation vectors learned by the pretrained OSFM based on limited homodyne measurement statistics. 
    Distinct clusters of optical states emerge, and largely reflect the ordering in mode number $m$, squeezing degree $\xi$, and stellar rank $r$.
    }
    \label{fig-4}
\end{figure*}
\subsubsection{Cat states}
Unlike the finite–stellar-rank families explored so far, we now consider Schr\"odinger cat states, a kind of states that have infinite stellar rank because their Fock support is unbounded. 
This places them much further beyond the non-Gaussian hierarchy explored in pretraining, offering a test of whether the model pretrained on finite–stellar-rank data extrapolates to the infinite–stellar-rank regime.
\par
A cat state is a superposition of two coherent states of equal amplitudes
\begin{align}
    \ket{\text{cat}}= \mathcal N_\alpha\left(\ket{\alpha}+e^{i\phi}\ket{-\alpha}\right),
\end{align}
where $\mathcal{N}_\alpha$ is a normalization factor, $|\alpha\rangle$ is a coherent state with amplitude $\alpha\in\mathbb C$, and $\phi$ is the relative phase. 
$|\alpha|^2$ is often referred to as the size of the cat.
These states exhibit interference fringes in their Wigner functions, creating negative regions in phase space, and serve as valuable resources in continuous-variable quantum computing~\cite{PhysRevA.68.042319,PRXQuantum.5.030322}, especially for quantum error correction~\cite{PhysRevA.59.2631,Grimm2020}.
The two-component cat code may be defined as $|+\rangle\propto|\alpha\rangle+|-\alpha\rangle$, $|-\rangle\propto|\alpha\rangle-|-\alpha\rangle$~\cite{Vlastakis2013-kv}.
However, this encoding method will introduce non-orthogonality, which depends on the overlap $|\langle\alpha|-\alpha\rangle|^2=e^{-4|\alpha|^2}$ related to the size of the state.
\par
In what follows, we show that the representation network of OSFM, pretrained on finite stellar-rank states, transfers effectively to the infinite-rank regime when inferring the size and fidelity of lossy cat states.
We generate $600$ random lossy cat states with $\alpha\in[0,2]$ and random phase $\phi\in[0,2\pi)$ for the task. 
While the cat states are of infinite stellar rank, in numerical experiments, we still represent them in a truncated Fock basis with a finite photon-number cutoff $N_{\mathrm{cut}}=24$.
Again, a lightweight property prediction network is attached to the representation network $\mathcal{E}$.
$10$ to $15$ marginals of each cat state are randomly selected as input. 
After fine-tuning, as shown in Fig.~\ref{fig-3}\textbf{c}, the coefficients of determination $R^2$ reach $0.9837$ for the size prediction and $0.9373$ for the fidelity prediction. 
In this case, the results highlight the effectiveness of our foundation model in extending to states with infinite stellar rank that are quite different from the pretraining dataset.

\subsubsection{Highly squeezed states}
Within the \((m,\xi,r)\)-space, after previously extending along the axes of mode number and stellar rank, a natural extension is then to push along the squeezing axis. 
Specifically, we investigate squeezed states up to \(\xi=10.4\,\mathrm{dB}\).
Squeezed states are crucial for advancing quantum metrology, which can enhance sensitivity in estimating unknown parameters, surpassing the standard quantum limit (SQL) imposed by classical resources~\cite{PhysRevLett.130.123603}.
While previously discussed N00N states are often referred to as the optimal state for loss-free sensing, they are extremely fragile to loss and lack deterministic phase estimation. 
By contrast, squeezed states with homodyne detection are deterministic and loss-tolerant, demonstrating great advantages in quantum metrology~\cite{4vdx-7224}.
\par
In quantum metrology, the quantum Fisher information (QFI) is a central quantity to estimate the achievable precision of the parameter from the quantum state.
For a unitary encoding $\rho_\theta = e^{-i\theta A}\,\rho\,e^{i\theta A}$ generated by an arbitrary observable $A$, the quantum Fisher information of $\rho$ is given by
\begin{equation}
F_Q[\rho,A]
= 2\sum_{k,l}\frac{(\lambda_k-\lambda_l)^2}{\lambda_k+\lambda_l}\,
\bigl|\langle k|A|l\rangle\bigr|^2 ,
\end{equation}
where $\{\lambda_k,|k\rangle\}$ are the eigenpairs of $\rho$ and the sum is over $k,l$ with $\lambda_k+\lambda_l>0$.
Here we construct the observable ${A}$ by optimizing over an arbitrary linear combination of the quadrature operators $x$ and $p$. 
\par
To evaluate whether OSFM can capture the complex nonlinear mapping from homodyne measurement data to the QFI of the states, $600$ lossy squeezed vacuum states are generated for fine-tuning.
For these Gaussian states, only $s=3$ marginals are used as input, corresponding to phases $\theta=\{0,\pi/4,\pi/2\}$. 
Fig.~\ref{fig-3}\textbf{d} reports the results.
OSFM exhibits excellent calibration, with predictions closely matching the ground truth.
The predicted QFI (left) and fidelity (right) tightly follow the identity line with coefficients of determination \(R^2=0.9846\) and \(R^2=0.9936\), respectively.
Residuals remain small across the full range, with only a slight spread at the largest QFI values. 

\subsection{Clustering of optical state families}
We further examine how these optical state families distribute in the representation space, together with the OOD states.
For each family, we randomly sample $500$ states and input their homodyne measurement marginals into the pretrained representation network $\mathcal{E}$ of OSFM. 
The representation vectors are then projected onto a 2D plane by t-SNE, visualized in Fig.~\ref{fig-4}.
Without any prior knowledge, distinct clusters of optical states emerge in the 2D embedding.
\par
Within the N00N state family, we consider photon numbers $1\leq N\leq8$, and the embedding naturally splits them into eight compact subclusters, which are plotted in detail in the Supplemental Material~\cite{supp}. 
Notably, the $N=1$ subcluster lies close to the multimode squeezed states with single-photon subtraction or addition, both of which share stellar rank $r=1$.
The $N=2$ subcluster lies close to the OOD states, on the side of lower stellar rank $r=3$.
The remaining subclusters are distributed following the ordering of $N$.
Cat states that have infinite stellar rank form an isolated island in the embedding space without any mixture with other families.
The single-mode squeezed states are located near the region of multimode squeezed states with mode-selected single-photon subtraction or addition, whose homodyne statistics partially resemble those of single-mode squeezed states. 
\par
Altogether, these clusters indicate that OSFM has learned a physically meaningful representation: it separates state families in a manner consistent with mode number $m$, squeezing degree $\xi$, and stellar rank $r$.

\section{Discussion}
 In this paper, we develop the first foundation model for the characterization of optical quantum states. By following the pretraining-fine-tuning paradigm, we show that this trained model can be successfully applied to a wide range of characterization tasks for optical quantum states, especially multimode non-Gaussian states.
 Specifically, through pretraining on homodyne measurement outcome statistics, the neural network model learns structural patterns of various continuous variable quantum states. These learned representations transfer robustly both to those states of higher complexity unseen during training and to those prediction tasks not encountered before. 
 This enables a single neural network to predict non-Gaussianity, quantum fidelity, metrological sensitivity, and other essential features directly from homodyne measurement data for a variety of experimentally relevant quantum states. Efficient characterization and certification of these optical quantum states, such as, NOON states and cat states, will enhance the development of optical quantum computation, communication and metrology. 
 
\par
Although our approach provides a unified framework for characterizing a wide range of optical quantum states, the resulting neural network remains lightweight and cannot be considered as a large general-purpose foundation model.
A more powerful foundation model should leverage versatile multimodal information to learn representations of optical quantum systems. For instance, it could utilize the graph structure of optical quantum circuits~\cite{flam2022learning,PhysRevLett.133.130601} to complement measurement data in learning the representations of optical output states. We leave this for future work.

\section{Methods}
\subsection{The stellar formalism}
To randomly generate $m$-mode optical quantum states for pretraining, we first characterize them using the stellar formalism~\cite{Chabaud2020stellar,Chabaud2021classical,Chabaud2022holomorphic} in the continuous-variable domain. 
In this formalism, we analyze an $m$-mode normalised pure state $|\psi\rangle$ in terms of its stellar function $F_\psi^{\star}(\mathbf{\alpha}) \equiv \text{e}^{\frac{1}{2}\|\mathbf{\alpha}\|^2}\left\langle \mathbf{\alpha}^*|\psi\right\rangle$, where $|\mathbf{\alpha}\rangle$ is a coherent state with complex amplitude $\mathbf{\alpha}$. 
The stellar rank $r$ of $|\psi\rangle$ is defined as the number of zeros of $F_\psi^{\star}(\mathbf{\alpha})$, representing a minimal non-Gaussian operational cost to engineer the state from the vacuum.
\par
By introducing the notation $\overline{\mathbb{N}} = \mathbb{N} \cup \{+\infty\}$, the stellar hierarchy of continuous-variable states is induced by the stellar rank $r\in\overline{\mathbb{N}}$.
For instance, $r=0$ means that the state is Gaussian, while $r=1$ corresponds to a class of non-Gaussian states that contains both single-photon-subtracted and -added states. 
Another important and experimentally accessible class of non-Gaussian states, cat states, are of infinite stellar rank $r\rightarrow\infty$.

\subsection{Optical states for pretraining}
Here we discuss the optical state generation procedure for pretraining in detail.
Any multimode pure state $|\psi\rangle$ with finite stellar rank $r$ can be decomposed into $|\psi\rangle= G|C\rangle$, where $ G$ is a Gaussian operator acting onto the state $|C\rangle$, which is called core state; it is a normalized pure quantum state with multivariate polynomial stellar function of degree $r$, equal to the stellar rank of $|C\rangle$. 
For example, the stellar function of a three-mode core state $|012\rangle$ is given by $\frac{1}{\sqrt{2}}z_2z_3^2$, whose degree is equal to the stellar rank $r=3$ of the state.
It then follows immediately that Gaussian operations $G$ must preserve the stellar rank~\cite{Chabaud2021classical}. \par
\par
Our randomly generated optical states follow the above decomposition, which begins with a core state $|C(r)\rangle$ with a given stellar rank $r$ and random complex superposition coefficients of the Fock basis. 
According to the Williamson decomposition and the Bloch-Messiah decomposition, an $m$-mode Gaussian unitary operation $G$ can be decomposed as $G=U(\varphi)\left( \prod\limits_{i=1}^m {S}_i(\xi_i){D}_i(\alpha_i)\right){V}(\phi)$, where ${ S}_i(\xi_i)$ is a single-mode squeezing operator with complex squeezing parameter $\xi_i$ acting on mode $i$, and ${ D}_i$ is a displacement operator with complex displacement amplitude $\alpha_i$ acting on mode $i$. 
$U(\varphi)$ and $V(\phi)$ are passive linear optical transformations over $m$ modes, consisting of $(m-1)$ beam splitters with complex coupling coefficients $\varphi$ and $\phi$, respectively.
\par
Optical losses are also added to each mode of the pure state $|\psi\rangle=G|C(r)\rangle$, using a single-mode loss channel $\mathcal{L}_i(\rho)=\sum\limits_{n=0}^\infty L_n^{(i)}\rho {L_n^{(i)}}^\dagger$, where each $L_i$ is the Kraus operator~\cite{ou1992realization,LossChannelLect,Eaton2022measurementbased} given by
\begin{align}
    L_n^{(i)}(\eta_i) = \sqrt{\frac{(1-\eta_i)^n}{n! \eta_i^n}} \hat{a}_i^n e^{\frac{1}{2}  \hat{a}_i^\dagger \hat{a}_i\text{ln}\eta_i},
\end{align} 
representing an $n$-photon loss with efficiency $\eta_i$. 
Here we truncate the degree at $n=10$.

\subsection{Dataset organization}
Here we discuss how the generated optical quantum state data are organized for different training and testing stages. 
For pretraining, we generate $6\,000$ states in a relatively small Hilbert space. 
To assess OOD generalization of the model, $500$ states from a larger Hilbert space are generated for testing.
For downstream transfer learning, we prepare separate datasets per state family, using comparatively few samples for fine-tuning to evaluate performance in the low-data regime. 
The task involving multimode single-photon-subtracted/added states uses $600$ training and $150$ test samples for each mode number $m\in{5,7,10}$.
For both single-mode highly squeezed states and cat states, we generate $1\,500$ samples per state family, splitting each dataset into $600$ for training and $900$ for testing.
For N00N states, we generate $2\,000$ samples, with $600$ training and $1\,400$ test samples. 

\subsection{OSFM framework}
In the pretraining phase, the OSFM consists of a representation network $\mathcal{E}$, a generation network $\mathcal{G}$. 
$\mathcal{E}$ encodes each pair of homodyne measurement settings and marginal distributions $\{M_j,P_j\}_{j=1}^s$ into a latent vector $z_j=\mathcal{E}(M_j,P_j)\in\mathbb{R}^{z_\mathrm{dim}}$, and the representation $z$ of the entire state is obtained by averaging over $s$ pairs. 
The generation network $\mathcal{G}$ takes $\mathbf{z}$ and an unseen measurement configuration $M^\prime$ as input and outputs the predicted probability distribution $P^\prime=\mathcal{G}(z,M^\prime)$. 
During pretraining, both $\mathcal{E}$ and $\mathcal{G}$ are optimized jointly with a composite loss with three terms: (i) a reconstruction likelihood between $P^\prime$ and $P_\text{true}$, (ii) KL regularization for seeking a better prior distribution of the hidden variable in the generation process, and (iii) triplet loss with a gradually increasing weight during training to shape the representation space. 
\par
After pretraining, the representation network $\mathcal{E}$ is reused and attached with a task-specific lightweight prediction network $\mathcal{D}$ for downstream fine-tuning. 
The prediction network $\mathcal{D}$ is a fully connected network with one or two hidden layers. 
The input to $\mathcal{D}$ is the state representation $r$ from pretrained $\mathcal{E}$.
Transfer learning in this stage fine-tunes all model parameters using mean-squared error (MSE) loss for regression and binary cross-entropy (BCE) loss for classification.
The model is trained using Adam optimizer and parallelized across available GPUs via PyTorch~\cite{NEURIPS2019_9015}.



\bibliography{references}



\newpage
\onecolumngrid
\appendix

\begin{figure*}[bht]
    \centering
    \includegraphics[width=\linewidth]{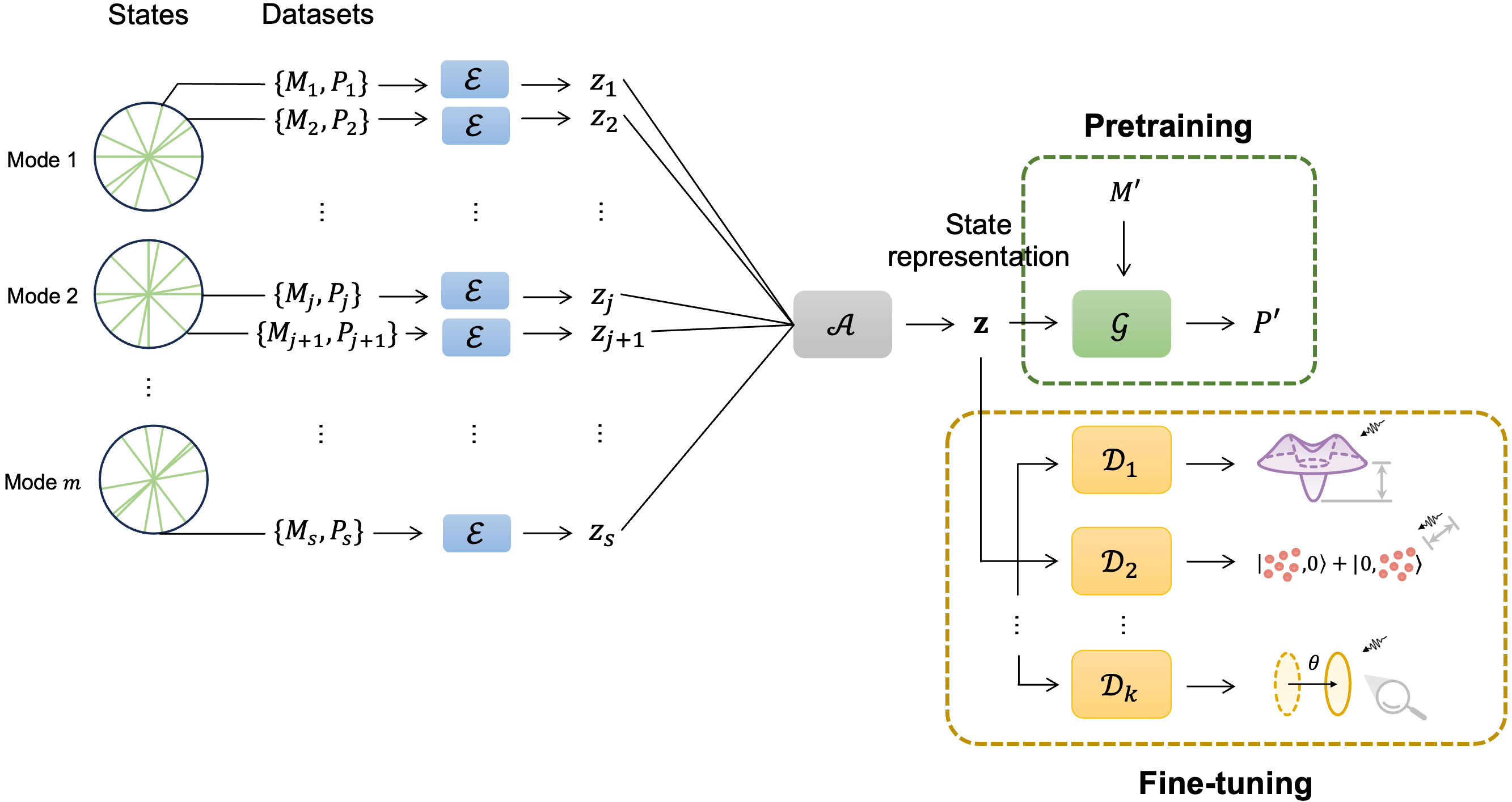}
    \caption{Structure of OSFM during pretraining and fine-tuning.}
    \label{fig:NN}
\end{figure*}

\section{Architecture of the networks}
Figure~\ref{fig:NN} illustrates the overall architecture of the foundation model for optical quantum state characterization (OSFM), which is designed to extract a unified state representation of multimode
optical quantum states from local homodyne measurements.
During pretraining, the model consists of a representation network $\mathcal{E}$, an aggregate function $\mathcal{A}$, and a generation network $\mathcal{G}$ that predicts quadrature marginals $P^\prime$ conditioned on unseen measurement setting $M^\prime$.
During fine-tuning, the averaged state representation from $\mathcal{A}$ is fed into different lightweight prediction heads $\{D_k\}$ for various downstream quantum property estimation tasks, with $k$ indexing specific target state families such as cat states and N00N states.
\par
\begin{figure*}
    \centering
    \includegraphics[width=\linewidth]{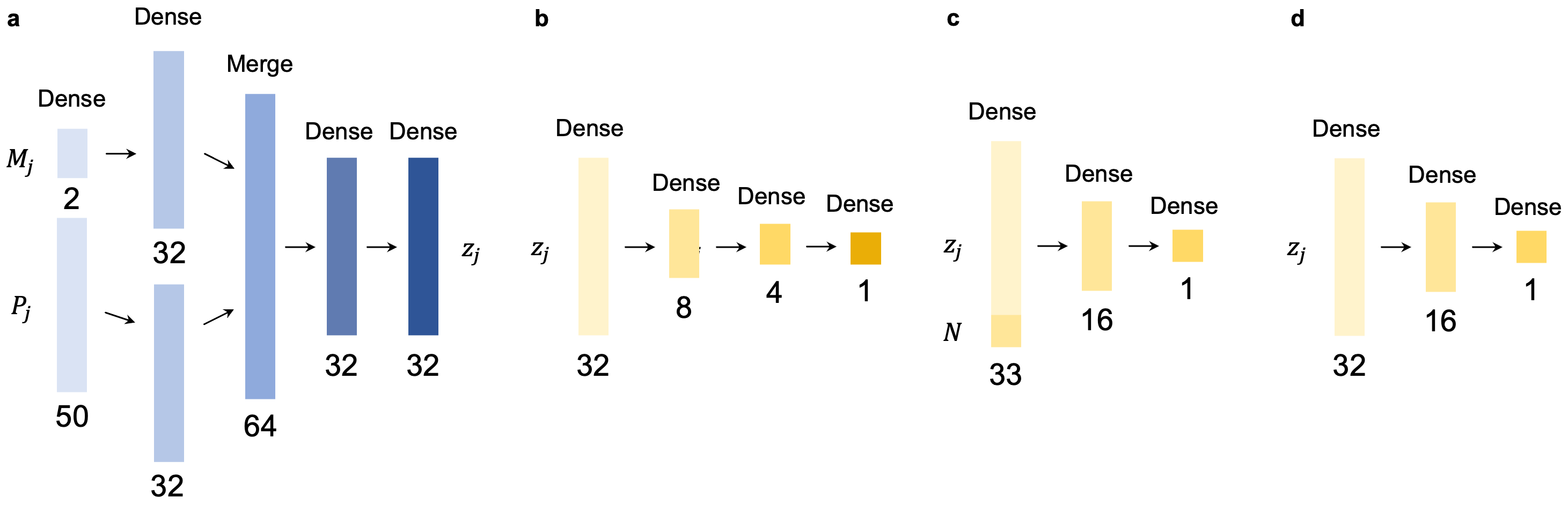}
    \caption{Architecture of the neural networks used in our model. \textbf{a}, Representation network $\mathcal{E}$ that encodes measurement setting $M_j$ and quadrature outcome $P_j$ into the state representation $z_j$. \textbf{b}, Prediction network used for estimating properties of multimode single-photon subtracted/added states. \textbf{c}, Prediction network for N00N states. \textbf{d}, Prediction network for cat states and highly squeezed states.
}
    \label{fig:NNArch}
\end{figure*}

The representation network $\mathcal{E}$ is a multi-layer perceptron (MLP) that encodes each measurement setting and outcome pair $\{M_j, P_j\}$ into an embedding $z_i \in \mathbb{R}^{z_{\mathrm{dim}}}$, as shown in Fig.~\ref{fig:NNArch}\textbf{a}.
Here, the measurement setting $M_j \in \mathbb{R}^2$ and the corresponding quadrature marginal $P_j \in \mathbb{R}^{50}$ are first processed independently by two dense layers, each outputting a 32-dimensional vector with ReLU activation. 
The outputs are concatenated into a 64-dimensional vector, which is further passed through two dense layers of size 32, again followed by ReLU activations, producing the final representation $z_j$. 
In our implementation, we set $z_{\mathrm{dim}} = 32$.
The aggregate function $\mathcal{A}$ computes the state representation $\mathbf{z} \in \mathbb{R}^{z_{\mathrm{dim}}}$ as the average of the set of measurement encodings ${z_j}$ for a given quantum state, which captures the overall state characteristics.
\par
In the fine-tuning phase, each prediction network $D_k$ is a MLP that takes $\mathbf{z}$ as input and outputs an estimate of a specific property or label of the quantum state. 
We design separate prediction networks tailored to each downstream task. 
For multimode single-photon subtracted/added states, the prediction head comprises three fully connected layers with hidden dimensions 8 and 4, followed by a output layer (Fig.~\ref{fig:NNArch}\textbf{b}). 
For N00N states, we include the photon number \(N\) as an additional input dimension, resulting in an input size of 33. This is processed by two dense layers with hidden dimension 16 and output dimension 1 (Fig.~\ref{fig:NNArch}\textbf{c}). For cat states and highly squeezed states, we only use the state representation as the input (Fig.~\ref{fig:NNArch}\textbf{d}).

\begin{figure*}[hbt]
    \centering
    \includegraphics[width=0.7\linewidth]{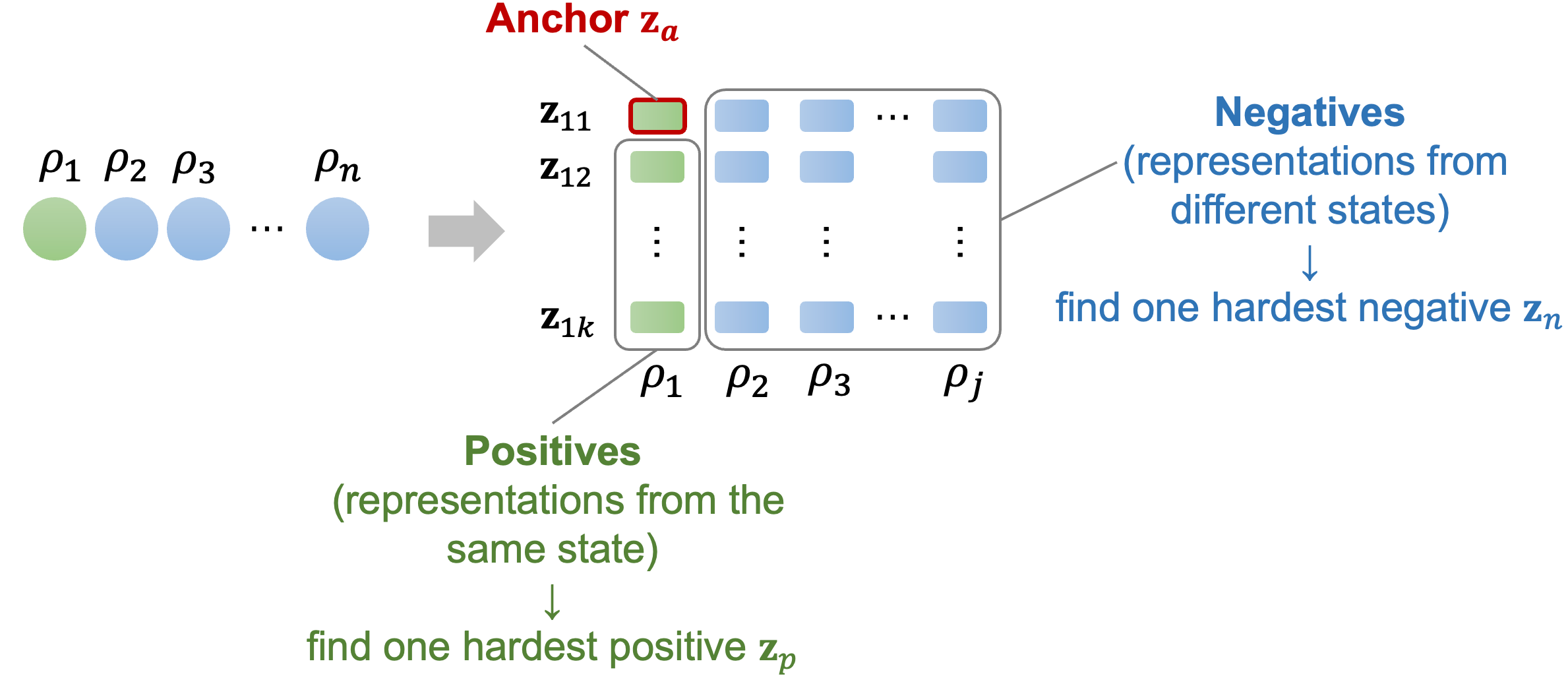}
    \caption{The triplet sampling strategy in a batch during pretraining.}
    \label{fig:triplet}
\end{figure*}

\section{Triplet loss implementation}
To enhance the structure of the learned latent representation space, we incorporate a triplet loss term during the pretraining phase. 
This additional loss encourages representations obtained from the same quantum state to cluster more tightly, while pushing apart those corresponding to different states. 
\par
As illustrated in Fig.~\ref{fig:triplet}, each training batch consists of $B$ quantum states, from which $K$ measurement subsets are sampled per state, resulting in $B \times K$ representations $\{ \mathbf{z}_{jk} \}$, where $j$ indexes the state and $k$ indexes the view. 
One representation $\mathbf{z}_a$ is randomly selected as the anchor. 
To construct a triplet, we identify the hardest positive $\mathbf{z}_p$ as the representation derived from the same state $j$ that is farthest from the anchor in terms of Euclidean distance, and the hardest negative $\mathbf{z}_n$ as the representation from a different state $j' \neq j$ that is closest to the anchor.
The Euclidean distances are computed as
$d(\mathbf{z}_a, \mathbf{z}_i) = \left\| \mathbf{z}_a - \mathbf{z}_i \right\|_2,$
where \( \mathbf{z}_a \) is the anchor and \( \mathbf{z}_i \) denotes a candidate positive or negative representation vector.
\par
The triplet loss is defined as
\begin{align}
\mathcal{L}_{\text{triplet}} = \max\left(0, \|\mathbf{z}_a - \mathbf{z}_p\|_2^2 - \|\mathbf{z}_a - \mathbf{z}_n\|_2^2 + \text{margin} \right),
\end{align}
where the margin is set to 1.0 in our implementation.
The triplet loss is combined with the loss which quantifies the discrepancy between the predicted and true quadrature distributions. 
A weighting coefficient $\lambda_{\mathrm{tri}}$ is linearly increased from 0 to 0.2 over the course of training.

\begin{figure*}
    \centering
    \includegraphics[width=0.9\linewidth]{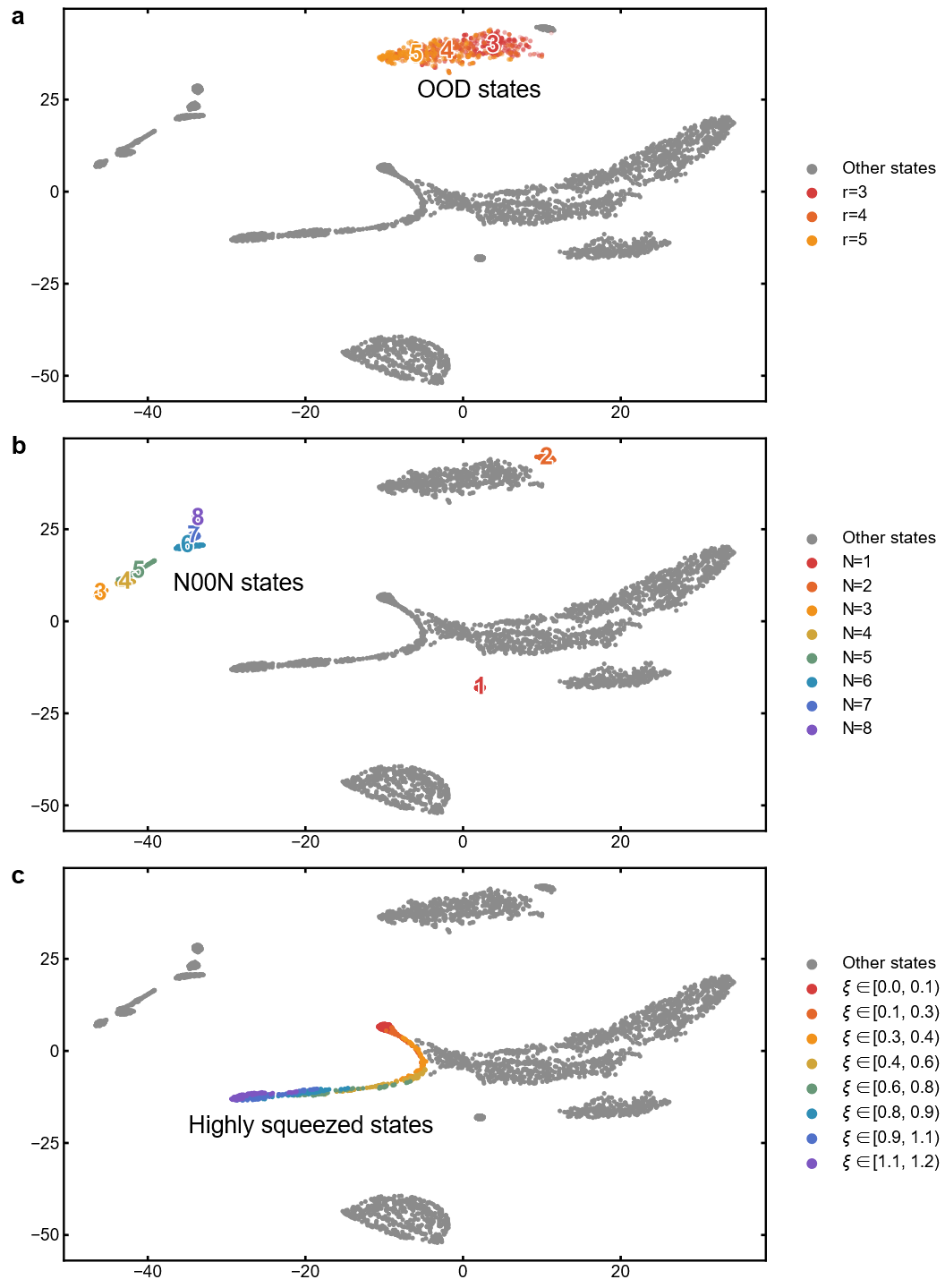}
    \caption{Visualization of OSFM representation space using t-SNE, highlighting subclusters corresponding to \textbf{a} out-of-distribution (OOD) states with stellar ranks $r=3,4,5$, \textbf{b} N00N states with varying photon number $N$, and \textbf{c} highly squeezed states with increasing squeezing parameter $\xi$.}
    \label{fig:subcluster}
\end{figure*}

\section{Subclusters within optical state families}
In the main text, we visualize how different optical state families cluster in the state representation space through the t-SNE algorithm.
To further examine the detailed structure learned by the OSFM model, we highlight the subclusters within specific optical state families. 
\par
In Fig.~\ref{fig:subcluster}\textbf{a}, out-of-distribution (OOD) states with stellar ranks $r = 3,4,5$ form a compact, isolated subcluster, demonstrating the model's ability to group unfamiliar but structurally related states. 
Panel \textbf{b} displays N00N states with photon numbers ranging from $N = 1$ to $N = 8$, which separate cleanly from the remaining data. 
The $N = 1$ subcluster lies close to the single-photon–added/subtracted squeezed states, and both correspond to stellar rank $r = 1$, indicating that the learned representations are sensitive to underlying rank structure even when Gaussian operations such as squeezing and beam-splitting are involved. 
Notably, although the $N = 1$ subcluster appears as a single dot, it actually contains 51 tightly packed states. 
In panel \textbf{c}, we visualize highly squeezed states with squeezing parameter \( \xi \in [0, 1.2] \). 
These states align along a smooth one-dimensional trajectory in latent space, reflecting a continuous progression in squeezing degree.
    
\end{document}